\documentclass[aps,amsfonts,nofootinbib,twocolumn]{revtex4}
\usepackage{amsmath}

\begin{document}

\newcommand{\bb}{\begin{equation}}
\newcommand{\ee}{\end{equation}}
\newcommand{\eqb}{\begin{eqnarray}}
\newcommand{\eqf}{\end{eqnarray}}

\preprint{}
\title{CPT/Lorentz Invariance Violation and Neutrino Oscillation }
\author{Paola Arias}
\email{paola.arias@gmail.com} \affiliation{Departamento
de F\'{\i}sica, Universidad de Santiago de Chile, Casilla 307,
Santiago 2, Chile}
\author{Ashok Das}
\email{das@pas.rochester.edu} \affiliation{Department of Physics and Astronomy,
University of
  Rochester, Rochester, NY 14627-0171, USA \\
  and
  \\
 Saha Institute of Nuclear Physics, 1/AF Bidhannagar, Calcutta 700064, India.}
\author{J.\ Gamboa}
\email{jgamboa@lauca.usach.cl} \affiliation{Departamento
de F\'{\i}sica, Universidad de Santiago de Chile, Casilla 307,
Santiago 2, Chile}
\author{J.\ L\'opez-Sarri\'on}
\email{justinux75@gmail.com} \affiliation{Department of
Physics, City College of CUNY, New York, NY 10031, USA}
\author{F.\ M\'endez}
\email{fmendez@lauca.usach.cl} \affiliation{Departamento de
F\'{\i}sica, Universidad de Santiago de Chile, Casilla 307, Santiago
2, Chile}

\begin{abstract}
We analyze the consequences of violation of Lorentz and $CPT$
invariance in the massless neutrino sector by deforming the canonical
anti-commutation relations for the fields. We show that, for
particular choices of the deformation, oscillation between massless
neutrino species takes place when only Lorentz invariance is
violated. On the other hand, if both Lorentz and $CPT$ invariances are
violated, we show that there is no oscillation between massless
neutrino species. Comparing with the existing experimental data on
neutrino oscillations, we obtain bounds on the parameter for Lorentz
invariance violation. 
\end{abstract}

\maketitle

\section{Introduction}

There has been an increased interest in the possibility that
Lorentz and $CPT$ symmetries may be violated at very high energies. For
example, recent developments in quantum gravity suggest that Lorentz
invariance may not be an exact symmetry at high energies \cite{amelino} and
$CPT$
invariance has also been questioned within such contexts \cite{kost2}.
Spontaneous
violation of $CPT$ and Lorentz symmetries can arise in string theories
\cite{kost3} and
the violation of Lorentz invariance in non-commutative field theories is
well known \cite{douglas}. On the experimental side, the UHE (ultra high energy)
cosmic ray events seen at AGASA \cite{agasa} and
presently under study by AUGER \cite{auger}
further support the possibility that Lorentz and $CPT$
invariances may not hold at such energies. Of course, there already exist
very stringent bounds on Lorentz and $CPT$ violation from laboratory
experiments in the Kaon and the lepton sectors and any violation of these
symmetries has to be compatible with these limits. Nonetheless, it is
possible that even a tiny violation of $CPT$ and Lorentz invariance can
lead to interesting mechanisms for physical phenomena. In a recent
paper, for example, we have shown \cite{us} how such a violation can lead to
baryogenesis
in thermal equilibrium (evading one of the criteria of Sakharov).
 In this note, we analyze the consequences of Lorentz and $CPT$
violation in the neutrino sector. We would like to emphasize that
several papers have already dealt with the effects of Lorentz
\cite{gla, bertolami} and $CPT$ violation in the neutrino sector,
particularly in connection with a qualitative discussion of neutrino
oscillation in this scenario \cite{cole1} (in another related
context see \cite{fogli}). In this paper, we carry out a
quantitative study of such phenomena within the context of a simple
model and derive bounds on such symmetry violating parameters from
the existing experimental results on neutrino oscillation.

Neutrino oscillation is an interesting phenomenon proposed about
fifty years ago by Pontecorvo (in a different context) which is
used to explain the deficit of solar and
atmospheric neutrinos in fluxes measured on earth
\cite{pontecorvo,w, sm} (for other recent analyses see  \cite{varios}). This
mechanism which is responsible for the
resolution of these puzzles is closely related to the $K^0$-${\bar K}^0$
oscillation \cite{pais}. In its simplest form, the probability for
oscillation between two species of particles $i,j$ with a mixing angle
$\theta_{ij}$ and energy levels $E_{i},E_{j}$ in a time interval $t$
is given by

\begin{equation}
P_{i\rightarrow j} (t) = \sin^{2} \left(2\theta_{ij}\right)\ \sin^{2}
\left(\frac{\Delta E_{ij} t}{2}\right),\label{osc}
\end{equation}
where
\begin{equation}
\label{ooo}
\Delta E_{ij} = E_{i}-E_{j}.
\end{equation}

If the oscillation is between two neutrino species $\nu_{i},\nu_{j}$
with small masses $m_{i},m_{j}$ respectively, then in the
conventional scenario one expands (this assumes Lorentz invariance
and $c=1$)
\bb
E_{i} = \sqrt{p^{2} + m_{i}^{2}} \approx p + \frac{m_{i}^{2}}{2p} =
p
+ \frac{m_{i}^{2}}{2E_{i}},
\ee
so that we have
\bb
\Delta E_{ij} = E_{i} - E_{j} \approx \frac{m^{2}_{i} - m^{2}_{j}}{2E}
= \frac{\Delta m_{ij}^{2}}{2E}, \label{osc2}
\ee
where we have assumed that for neutrinos of small mass, $E_{i}\approx
E_{j} = E$.
In this case, the probability for oscillation between the two neutrino
species in traversing a path length $L$ can be written as (see
\eqref{osc} and \eqref{osc2})
\begin{eqnarray}
P_{\nu_i\rightarrow\nu_j} (L) & = & \sin^{2}
\left(2\theta_{ij}\right) \sin^{2}\left(\frac{\Delta m_{ij}^{2}
    L}{4E}\right)\nonumber\\
& = &
\sin^2\left(2\theta_{ij}\right)~\sin^2\left(\frac{1.27\Delta
    m^{2}_{ij} L}{E}\right), \label{osc1}
\end{eqnarray}
where $\Delta m^{2}_{ij} = m_{i}^{2} - m_{j}^{2}$ is taken in $({\rm
  eV})^{2}$, the neutrino energy $E$ in MeV and the length of path
traversed in `m' (meters) (In the last line of the above formula,
we have restored all the nontrivial constants as well as traded the
time interval for the path length assuming that the neutrino travels
almost at the speed of light.)

It follows from eq.(\ref{osc1}) that neutrino oscillation does
not take place in free space if neutrinos are massless or (when
massive) are degenerate in mass. With three families of neutrinos,
there can only be two independent combinations of squared mass
differences, say $\Delta m^{2}_{12}, \Delta
 m^{2}_{23}$ which are sufficient to find a solution for the solar
neutrino as well as the atmospheric neutrino puzzles. Within the
standard model, this can be achieved with the bounds \cite{exp}
\bb
\Delta m^{2}_{12} \leq 10^{-4} {\rm eV}^{2},\quad 10^{-3} {\rm
  eV}^{2}\leq \Delta m^{2}_{23}\leq 10^{-2} {\rm eV}^{2}.
\label{bounds}
\ee
Given these, the bound on $\Delta m^{2}_{13} = \Delta m^{2}_{12} +
\Delta m^{2}_{23}$ is determined. There is no further freedom within a model
with
three families of neutrinos.

Several experiments by now have looked for neutrino oscillations. One
such experiment, namely, the LSND (Liquid Scintillator Neutrino
Detector at Los Alamos) \cite{LSND} has used muon sources from the
decay $\pi^{+}\rightarrow \mu^{+} + \nu_{\mu}$. The experiment looks
for neutrino oscillation in the subsequent decay of the muon through
$\mu^{+}\rightarrow e^{+} + \nu_{e} + \bar{\nu}_{\mu}$. After a
path length of $L=30$m, the experiment finds the oscillation channel
$\bar{\nu}_{\mu}\rightarrow \bar{\nu}_{e}$ (with $20 {\rm MeV}\leq
E_{\nu_{\mu}} \leq 58.2 {\rm MeV}$) with a probability of
$0.26$\%. The experiment also reports the existence of the oscillation
$\nu_{\mu}\rightarrow \nu_{e}$ with the same probability. Furthermore,
the analysis of the results of this 
experiment, following \eqref{osc1},
leads to a bound on the difference of the relevant squared mass difference to be
\bb
\Delta m^{2} < 1 {\rm eV}^{2}.\label{lsnd}
\ee
However, the MiniBoone experiement, which was expected to verify the
results of LSND, has recently reported their first result
\cite{miniboone} and  excludes the  mass region in \eqref{lsnd}. As a
result, the simple explanation of the LSND results, based on the
two flavor neutrino oscillation is ruled out. 


All of the above discussion has been within the context of the
standard model with a massive neutrino where both Lorentz invariance and $CPT$ are assumed to
hold. On the other hand, if Lorentz invariance or $CPT$ or both are
violated
in the neutrino sector, it has been suggested that neutrino oscillation can take place in
free space even for massless neutrinos (in contrast to
eq. (\ref{osc1}) where Lorentz invariance is assumed). This was
pointed out by Coleman and Glashow
\cite{coleman} and developed more extensively by Kostelecky and
collaborators \cite{kost,green}. This is particularly clear from
eq. (\ref{osc}) where we see that the probability of oscillation
really depends on the difference in the energy of the two neutrino
species and if $\Delta E_{ij} = E_{i}-E_{j} \neq 0$ even when the
masses vanish, the probability of oscillation will be nontrivial. This
can happen, for example, if the two neutrino species have different
(energy) dispersion relations. This possibility has been discussed
extensively in the last few years by various groups
\cite{kost,barenboim}. In particular, ref \cite{kost} analyzes the
structure of the most general Lagrangian with violations of Lorentz
invariance and $CPT$ in an attempt to understand the discrepancy
between solar and atmospheric neutrinos and the LSND anomaly
\cite{LSND}.

The goal of this note is to analyze the consequences of violations of
Lorentz and $CPT$ invariances in the massless neutrino sector from the
point of view of a non-commutative field theory where such violations
are more natural. In such a model, the violation of Lorentz and $CPT$
invariances is  implemented through a deformation of the canonical
anti-commutation relations for the neutrino fields. Such a model can
be thought of as a subclass of the general model proposed in
\cite{kost}, but since, depending on the deformation, we have fewer
arbitrary parameters, we naturally have more predictive power. The
result of our analysis can be summarized as follows. If there is
violation of only Lorentz invariance, then oscillations between
massless neutrino species can take place and comparing with the
existing experimental data, we can determine bounds on the parameter
characterizing Lorentz invariance violation. On the other hand, if
both Lorentz and $CPT$ invariances are violated, there is no
oscillation between massless neutrino species.

\section{The Model and the Phenomenology of Neutrino Oscillation}

The model that we will describe below is inspired by the quantum
theory of non-commutative fields developed in \cite{nos}. The
quantum theory of fermionic non-commutative fields (neutrinos) is
obtained from
the standard fermionic quantum field theory by deforming the
anti-commutation relations while retaining the usual Hamiltonian.
In order to
explain in some detail the construction, let us consider the
conventional Lagrangian density for two flavors of massless fermions (neutrinos)
given
by
\bb
{\cal L} = i {\bar \psi}^{i} \gamma^\mu \partial_\mu \psi^i,
\label{la1}
\ee
where the superscript $i=\{1,2\}$ runs over the flavor quantum number (sum over
repeated indices is understood).

The Hamiltonian density has the form
\bb
{\cal H} = -i\left( \psi^{i \dagger} {\vec \alpha} \cdot~ \vec{\nabla} \psi^{i}
\right), \label{dh1}
\ee
where $\vec{\alpha} = \gamma^{0}\vec{\gamma}$. With the conventional canonical anti-commutation relations for the
fermion fields, one would obtain the standard relativistic equations for the
massless neutrinos using the Hamiltonian following from
\eqref{dh1}. However, the non-commutative theory is obtained by
deforming the canonical anti-commutation relations while maintaining
the form of the Hamiltonian density \eqref{dh1}.

We postulate the deformed equal-time anti-commutation relations to have the form (with all others
vanishing)
\bb
\{ \psi_\alpha^i ({\bf x}),\psi_\beta^{j\dagger} ({\bf y})\} = {\cal A}_{\alpha \beta }^{ij}
~\delta^{(3)} ({\bf x} - {\bf y}), \label{ant1}
\ee
where $\alpha, \beta,...=1,2,3,4$ are spinor indices and ${\cal
  A}_{\alpha \beta }^{ij}$ is a constant matrix. In this paper we
consider the following two special choices for the structure of the
deformation matrix ${\cal A}$. 
\begin{enumerate}
\item  ${\cal A}$ has a nontrivial structure only in the flavor space.

\item  ${\cal A}$ depends on both the flavor and the spinor indices nontrivially through a constant background vector.
\end{enumerate}
As we will see, the first choice leads to a violation of Lorentz
invariance whereas both Lorentz and $CPT$ invariances are violated
with the second choice. 

\subsubsection{Deformation depending only on flavor indices}

In this case, ${\cal A}$ is a $2\times 2$ constant matrix
with complex elements in general, which, for simplicity,  can be chosen to have the form
\bb
{\cal A}^{ij} = \left(\begin{array}{cc} 1 & \alpha \\
\alpha^\ast & 1 \end{array}\right),\label{amatri}
\ee
so that the complex parameters $\alpha$ can be thought of as the parameters of
deformation. Clearly, the deformed anti-commutation relations reduce to
the conventional ones when the parameters of deformation vanish.

Given the deformed anti-commutation relations (\ref{ant1}) and the Hamiltonian
density \eqref{dh1}, the dynamical equation takes the form
\bb
{\dot \psi}^i = -{\cal A}^{ij} \left( {\vec \alpha}\cdot \vec{\nabla}
\psi^j\right),
\ee
which in momentum space takes the form
\bb
E\psi^i= {\cal A}^{ij} \left( {\vec \alpha}\cdot {\vec p}\ \psi^j\right).
\label{mom1}
\ee
In order to determine the energy eigenvalues for this system, let us consider the unitary matrix $D$
\begin{eqnarray}
D &= & \frac{1}{\sqrt{2}} \left(\begin{array}{cc} \frac{|\alpha|}{\alpha}
  & 1 \\
-\frac{|\alpha|}{\alpha} &1\end{array}\right),\nonumber\\
D^{\dagger} &=& D^{-1} = \frac{1}{\sqrt{2}} \left(\begin{array}{cc}
  \frac{\alpha}{|\alpha|} & -\frac{\alpha}{|\alpha|} \\
1 &1\end{array}\right). \label{change}
\end{eqnarray}
It is straightforward to check that $D$ diagonalizes ${\cal A}$ and as a result,
the energy spectrum for the fermions follows to be ($c=1$)
\eqb
E^{1}_\pm &=& \pm \left( 1 +|\alpha| \right) |{\vec p}|,
\nonumber
\\
E^{2}_\pm &=& \pm \left( 1 -|\alpha| \right) |{\vec
  p}|.
 \label{disp2}
\eqf
Here $E^{1,2}$ are the energies of the two species of the neutrinos considered. We, therefore,
conclude that with this choice of the deformation, this system
exhibits violation of  Lorentz invariance,  as can be seen from the
dispersion relations (\ref{disp2}). However,
$CPT$ symmetry remains intact in this case which can be easily seen as follows.

As we have emphasized, our deformation can be thought of as a subclass
of  the standard model extension (SME)  \cite{kost2}. Indeed, if we
restrict to the part of the Lagrangian in  \cite{kost2}  given by 
\bb
{\cal L} = i {\bar \psi}^{i} \gamma^\mu \partial_\mu \psi^i + i {\bar \psi}^i c^{\mu \nu, ij} \gamma_\mu \partial_\nu \psi^j,
\ee
with the constant background field $c^{\mu\nu, ij}$ diagonal in the
space-time indices, then we obtain  $\pi^i= i( \delta^{ij} + c^{00,
  ij} )\psi^{j\dagger}$. This leads to the canonical anti-commutation
relations
\eqb
\{\psi^i ({\bf x}), \psi^{j\dagger} ({\bf y})\} &=& ( \delta^{ij} + c^{00, ij} )^{-1} \delta ({\bf x}-{\bf y}), \nonumber
\\
&=& {\cal A}^{ij} \delta ({\bf x}-{\bf y}),
\eqf
which is in agreement with our deformation for the non-commutative fields.
In this case, since the background field is a constant second rank tensor, the extra term
\[
{\bar \psi}^i c^{\mu \nu}_{ij} \gamma_\mu \partial_\nu \psi^j,
\]
violates Lorentz invariance but not $CPT$ symmetry as was also pointed out in reference \cite{kost2}.

The energy eigenstates corresponding to the eigenvalues \eqref{disp2} can now be
determined directly through the application of the diagonalizing matrix
\bb
D \left(\begin{array}{c} \psi^1 \\ \psi^2 \end{array}\right) =
\left(\begin{array}{c} {\tilde \psi}^1 \\ {\tilde
    \psi}^2\end{array}\right), \label{diagonal}
\ee
where $\tilde{\psi}^{1}$ and $\tilde{\psi}^{2}$ are eigenstates
with energy values $E^{1}$ and $E^{2}$ respectively.

The time evolution for the energy eigenstates, is determined to be
\eqb
{\tilde \psi}^1 (t) &=& e^{-i E_+^{1} t + i {\vec p}\cdot{\vec x}}
~{\tilde
  \psi}^1(0),
\\
{\tilde \psi}^2 (t) &=& e^{-i E_+^{2} t + i {\vec p}\cdot{\vec x}}~{\tilde
\psi}^2 (0),
\eqf
where $E^{1,2}_+$ are the energy eigenvalues determined in
(\ref{disp2}).
The diagonal wavefunctions can be seen from
\eqref{change} and
\eqref{diagonal} to have the explicit forms
\eqb
{\tilde \psi}^1&=& \frac{1}{\sqrt{2}} \left( \frac{|\alpha|}{\alpha}
\psi^1 + \psi^2\right), \nonumber
\\
{\tilde \psi}^2&=& \frac{1}{\sqrt{2}} \left(-
\frac{|\alpha|}{\alpha} \psi^1 + \psi^2\right)\cdot \label{tr2}
\eqf
Thus, we see that if $\alpha \in \Re$, then the diagonal
wavefunctions can be thought of as resulting from a rotation by
$\pi/4$ in the flavor space. This is, in fact, consistent with the
hypothesis of
large mixing angle (LMA) \cite{LMA} and, therefore, for simplicity let us choose
$\alpha$ to be real. In this case, we can parametrize (\ref{tr2}) as
\eqb
{\tilde \psi}^1&=& \cos\theta_{12} \psi^1 + \sin\theta_{12}
\psi^2, \label{tr11}
\\
{\tilde \psi}^2&=& -\sin\theta_{12} \psi^1+ \cos\theta_{12}
\psi^2, \label{tr22}
\eqf
with the mixing angle $\theta_{12} = 45^{\circ}$.

Relations (\ref{tr11}) and (\ref{tr22}) can now be inverted to give
\begin{eqnarray}
\psi^{1} &=& \cos \theta_{12} \tilde{\psi}^{1} -
\sin\theta_{12} \tilde{\psi}^{2},\nonumber\\
\psi^{2} & = & \sin\theta_{12} \tilde{\psi}^{1} +
\cos\theta_{12} \tilde{\psi}^{2}.
\end{eqnarray}
Thus, a neutrino initially in the state $\psi^{1}$ would evolve in time as
\eqb
\psi^1 (t) &=& \cos \theta_{12} ~{\tilde \psi}^1 (t) -
\sin \theta_{12} ~{\tilde \psi}^2 (t) \nonumber
\\
&=& \biggl[ \left( \cos^2 \theta_{12}~e^{-iE_+^{1} t} +
  \sin ^2\theta_{12}~ e^{-iE_+^{2} t}\right) \psi^1(0)
\nonumber
\\
&+& \frac{1}{2} \sin2 \theta_{12} \left( e^{-iE_+^{1} t} - e^{-iE_+^{2}
    t}\right) \psi^2 (0)\biggr]e^{i{\vec p}\cdot{\vec x}} \nonumber
\eqf
Therefore, at a later time $t$, the probability of finding the state
$\psi^2$ in the beam is given by
\eqb
P_{\nu_1 \rightarrow \nu_2} &=& \biggl \vert \frac{1}{2} \sin
2\theta_{12}~\left(
e^{-iE_+^{1} t} -
e^{-iE_+^{2} t}\right) e^{i{\vec p}\cdot{\vec x}}\biggr\vert^2, \nonumber
\\
&=&\sin^2 \left(2\theta_{12}\right)~ \sin^2 \left(\vert \alpha \vert \vert {\vec
  p}\vert t\right), \label{prob2}
\eqf
and since we are considering particles with velocities close to $c$, we
can replace
\bb
t\rightarrow L, \nonumber
\ee
where $L$ denotes the path length traversed by the neutrino. Let us note here that in our theory, the velocity of the neutrino
can in principle be different from $c$, but any further correction is suppressed by terms of the order ${\cal O}(\alpha^2)$ which is
extremely small. Thus, the probability for oscillation \eqref{prob2} becomes
\bb
P_{\nu_1 \rightarrow \nu_2}= \sin^2 \left(2\theta_{12} \right) ~ \sin^2
\left( \vert \alpha \vert~E ~L\right),\label{prob3}
\ee
where we have used the fact that for $|\alpha|\ll 1, E\approx |\vec{p}|$.

There are several things to note from the expression
\eqref{prob3}. First, the deformation parameter $\alpha$ leads to
rotations in the flavor space and thereby determines the mixing
angles. However,  differences from the conventional description of
massive neutrinos arise because these deformation parameters also
determine the nontrivial dispersion relations for the energy
eigenvalues and lead to a nontrivial energy difference even in the
absence of masses. Consequently, oscillation takes place even for
massless neutrinos. The difference from the conventional description
of neutrino oscillation shows up in 
\eqref{prob3} in the fact that the energy dependence is linear as
opposed to the inverse dependence in \eqref{osc1}.

\subsubsection{Vector-dependent deformations}

 In order to define the minimal matrix ${\cal A}$ which breaks both
$CPT$  and Lorentz symmetries and which connects different flavors --that is
${\cal A}^{12}\neq 0$ for two flavor indices -- we need to include a
constant background (real) vector $e_\mu^{ij}$. With this, the minimal ${\cal A}$ has the form
\bb
 {[{\cal A}]}_{\alpha \beta}^{ij} = \delta_{\alpha \beta} \delta^{ij} + (\gamma^\mu)_{\alpha \beta}
e^{ij}_{\mu}. \label{cho}
\ee
If we assume rotational invariance, we can set the space components of $e_{\mu}^{ij}$ to zero.
The simplest case mixing flavors would then correspond to choosing
$e_{0}^{12} = e = e_{0}^{21}$ and the equations of motion in this case
would have the forms 
\begin{eqnarray}
\label{veceq}
\dot{\psi}^1&=&- \vec{\alpha}\cdot\vec{\nabla}\psi^1 -
 e\vec{\gamma}\cdot\vec{\nabla}\psi^2, \nonumber
\\
\dot{\psi}^2&=&- \vec{\alpha}\cdot\vec{\nabla}\psi^2 - e\vec{\gamma}\cdot\vec{\nabla}\psi^1.
\end{eqnarray}
These can be diagonalized and lead to the dispersion relations
\eqb
E^{1}_\pm &=& \pm \sqrt{1+e^2}\ |{\vec p}|,
\nonumber
\\
E^{2}_\pm &=& \pm \sqrt{1+e^2}\ |{\vec p}|.
 \label{disp3}
\eqf
This is identical to the conventional dispersion relation for massless
neutrinos, expect for a scale factor, and shows in particular that
massless neutrinos and  anti-neutrinos of different species are
degenerate in energy. As a consequence, oscillations between massless
neutrino species cannot take place (see eq. (\ref{ooo})) if Lorentz
and $CPT$ symmetries are simultaneously violated in this model.

Equations (\ref{veceq}) can be also obtained from the Hamiltonian density
\bb
H=-i{\psi^i}^\dag\vec{\alpha}\cdot\vec{\nabla}\psi^i -i e_{0}^{ij}
{\psi^i}^\dag\vec{\gamma}\cdot\vec{\nabla}\psi^j, \label{haa}
\ee
with canonical anti-conmutation relations instead of (\ref{ant1}). Our
deformation can also be understood as a subclass of the 
standard model extension of \cite{kost2} as follows. Let us consider the part of
the extended model \cite{kost2} of the form 
\bb
{\cal L} = i {\bar \psi}^{i} \gamma^\mu \partial_\mu \psi^i + i {\bar \psi}^i
e^\mu_{ij} \partial_\mu \psi^j, \label{modi}
\ee
with our minimal choice for the constant background vector
$e_{\mu}^{ij}$. In this case, the derivation of the canonical
anti-commutation relation leads to the deformation discussed above up
to a field rescaling. Since the constant background field is a vector,
the extra term in the Lagrangian violates $CPT$ invariance and the
Lorentz violation of this term is manifest as well. Thus, the second
choice of the deformation violates both Lorentz and $CPT$
invariances. The $CPT$ violation can also be seen from the analysis of
the energy eigenstates. The diagonalized fermions involve a
combination of left and right handed fields and, as a consequence, do
not have well defined $CPT$ transformation properties. 

\subsubsection{Bounds for $\alpha$ (Lorentz invariance violation)}

 As we have shown, oscillation between massless neutrinos can take
 place in our model if only Lorentz invariance is violated. Therefore,
 let us use the existing experimental data on neutrino oscillations to
 derive bounds on the parameter $\alpha$ of Lorentz invariance
 violation (LIV). We note that our earlier analysis for two neutrino
 flavors can be extended to incorporate more flavors easily. For
 example, to accommodate three neutrino flavors, we need to generalize
 the deformation parameter (as well as the mixing angles) as 
\begin{equation}
\alpha\rightarrow \alpha_{ij},\quad
\theta_{12}\rightarrow \theta_{ij},
\end{equation}
where $i,j=1,2,3$. In this case, the probability for oscillation
between neutrino flavors can be written as
\begin{equation}
P_{\nu_{i}\rightarrow \nu_{j}} (L) = \sin^{2} \left(2\theta_{ij}\right) \sin^{2}
\left(|\alpha_{ij}| EL\right).\label{probij}
\end{equation}
The same formula also holds for anti-neutrino oscillations since $CPT$ is not violated in this case.

Therefore, comparing with the conventional analysis of oscillation for
massive neutrinos given in \eqref{osc1}, we can identify ($c=1$) 
\begin{equation}
|\alpha_{ij}| = \frac{\Delta m_{ij}^{2}}{4E^2}.\label{conversion}
\end{equation}
We note here that there are three deformation parameters
$\alpha_{ij},$
without any further constraint unlike the constraint on the
difference of the squared masses in the conventional scenario.

Let us next note that from the solar neutrino experiments, we know
that this involves oscillations of the flavors $1\rightarrow 2$ with
\begin{equation}
\Delta m_{12}^{2} < 8\times 10^{-5} {\rm eV}^{2}, \quad E\sim 1 {\rm
  MeV}.
\end{equation}
From \eqref{conversion}, this translates into a deformation parameter
\begin{equation}
|\alpha_{12}| < 10^{-17}.\label{result1}
\end{equation}
The atmospheric neutrino results, on the other hand, involve an
oscillation of the type $2\rightarrow 3$ with
\begin{equation}
\Delta m_{23}^{2} < 2.6\times 10^{-3} {\rm eV}^{2},\quad E\sim 1
{\rm GeV}.
\end{equation}
From \eqref{conversion}, we see that this would translate into a
deformation parameter
\begin{equation}
|\alpha_{23}| < 10^{-22}.\label{result2}
\end{equation}

Finally, we note that although the simple interpretation of the LSND
results has been disproved by the MiniBooNe 
experiment, it is nonetheless interesting to recognize that here the
oscillations involve flavors of the type $1\rightarrow 2$ (both in the
neutrino as well as the anti-neutrino channels) with 
\begin{equation}
\Delta m_{12}^{2} < 1 {\rm eV}^{2},\quad E \sim 50 {\rm MeV}.
\end{equation}
In this case, the analog of \eqref{conversion} for the antineutrinos
leads to
\begin{equation}
|\alpha_{12}| < 10^{-16}.\label{result3}
\end{equation}
It is clear now that within this scenario, all the experimental results
can be naturally explained without any particular puzzle. 
 We would like to note here that
our analysis for solar neutrinos are not in contradiction with the
data from the KamLAND experiment \cite{KAM}.

In this discussion, we have assumed the neutrinos to be completely
massless in which case, the conventional oscillation does not take
place. It is possible that the neutrinos have a small mass and that
both mechanisms do contribute to the phenomena of neutrino
oscillation. In this case, a careful analysis of the atmospheric
neutrino oscillation results can lead to even a more stringent bound
on the parameter $\alpha_{23}$ \cite{fogli}.

\section{Conclusion and outlook}

In this paper, we have carried out a quantitative analysis of the
consequences of $CPT$ and
Lorentz invariance violation in the massless neutrino sector. While it has
already been suggested that in such a case,
neutrino oscillation can take place even for massless neutrinos, we
have presented a simple model of
a theory of noncommutative fermions to study this phenomenon
quantitatively. The model contains a minimal number of symmetry
violating parameters that are introduced as deformation parameters in
the equal-time anti-commutation relations for the fermion
fields. Real values of these deformation parameters naturally lead to
the large mixing angle scenario. While the deformation parameters
directly lead to mixing between
different neutrino flavors, they also lead to nonstandard (energy)
dispersion relations (through Lorentz and $CPT$ violation), which
leads to oscillations between massless neutrino species if only
Lorentz invariance is violated. 

In the case that there is violation of only Lorentz invariance, we
have determined bounds on the parameters of deformation (parameters
characterizing Lorentz invariance violation) from the existing 
experimental data on solar and atmospheric neutrinos  as well as from
the LSND data. 
The bounds on the
deformation parameters within this minimal model are obtained to have
the values
\begin{equation}
\label{results}
|\alpha_{23}|< 10^{-22},\,\,\,\,\,\,\,|\alpha_{12}|< 10^{-17}
\end{equation}
We note that bounds for $\alpha$ were obtained from the LMA scenario
by using $\theta\sim\pi/4$. For the solar and atmospheric neutrinos, this is
indeed consistent with the experimental determination.

On the other hand, if Lorentz and $CPT$ invariances are violated simultaneously, then oscillation between massless neutrinos disappears in our model, although oscillations can take place for massive neutrinos in the standard scenario. This fact could indicate that  at very high energy where masses can be neglected, neutrino oscillation would signal a violation of Lorentz invariance and not of   $CPT$ symmetry.

\bigskip

\noindent{\bf Acknowledgment:}
This work was supported in part by US DOE Grant number DE-FG-02-91ER40685 and
by FONDECYT-Chile grants 1050114, 1060079, 3060002 and 2105016. J.L-S was
supported by a Fulbright Fellowship.

\end{document}